# Combining (Non-) Linear Optical and Fluorescence Analysis of DiD to Enhance Lipid Phase Recognition

Silvio Osella,[1,2] Florent Di Meo,[3] N. Arul Murugan,[2] Gabin Fabre[4], Marcel Ameloot,[5] Patrick Trouillas,[3,6] Stefan Knippenberg[2,5*]

1. Centre of New Technologies, University of Warsaw, Banacha 2C, 02-097 Warsaw, Poland.
2. Department of Theoretical Chemistry and Biology, School of Engineering Sciences in Chemistry, Biotechnology and Health, Royal Institute of Technology, SE-10691 Stockholm, Sweden.
3. UMR 1248 INSERM, Limoges University, Faculty of Pharmacy, 2 rue du Docteur Marcland, 87025 Limoges Cedex, France.
4. LCSN-EA1069, Faculty of Pharmacy, Limoges University, 2, rue du Dr. Marcland, 87025 Limoges Cedex, France
5. Biomedical Research Institute, Hasselt University and Transnational University Limburg, B-3590, Diepenbeek, Belgium.
6. Centre of Advanced Technologies and Materials, Faculty of Science, Palacký University, tř. 17 listopadu 12, 771 46 Olomouc, Czech Republic.

AUTHOR EMAIL ADDRESSES: sknippen@kth.se (SK)

**Abstract**

The widespread interest in phase recognition of lipid membranes has led to the use of different optical techniques to enable differentiation of healthy and not fully functional cells. In this work, we show how the combination of different (non-)linear optical methods such as One Photon Absorption (OPA), Two Photon Absorption (TPA) and Second Harmonic Generation (SHG) as well as the study of the fluorescence decay time leads to an enhanced screening of membrane phases using a fluorescent 1,1'-dioctadecyl-3,3,3',3'-tetramethylindocarbocyanine (DiD) probe. In the current study we consider the pure liquid disordered phases of DOPC (Dioleoyl-sn-glycero-3-phosphocholine, room temperature) and DPPC (1,2-Dipalmitoyl-sn-glycero-3-phosphocholine, 323 K), the solid gel phase of DPPC (298 K), and the liquid ordered phase of a 2:1 binary mixture of Sphingomyelin and Cholesterol. By means of extensive hybrid quantum mechanics – molecular mechanics calculations and based upon the (non-) linear absorption of the embedded probes, it is found that DiD can be used to identify the lipid bilayer phase. The joint TPA and SHG as well as fluorescence analyses qualifies DiD as versatile probe for phase recognition. In particular, the SHG data obtained by means of Hyper-Rayleigh Scattering and by Electric Field Induced Second Harmonic Generation reveal differences in polarization of the probe in the different environments. The TPA results finally confirm the particular location of the probe in between the polar head group region of the 2:1 SM:Chol mixture in the liquid ordered phase.

KEYWORDS Phase Recognition, (Non)-Linear Optical Properties, Fluorescence Lifetime, Hybrid QM/MM.

# INTRODUCTION

The use of fluorescent probes in investigations of lipid membrane phases has become a widely spread application. In particular, fluorescence spectroscopies, microscopies and fluorescence lifetime imaging (FLIM) techniques[1,2,3,4] allow differentiation between the different phases of lipid membranes. This has boosted the research for probes with enhanced photophysical properties.[5,6,7,8,9] This screening is of high importance in bio-medicine, since membrane phases can strongly affect cell functioning. In particular, it has been reported that cancer cells have increased membrane fluidity and polarity compared to healthy cells.[10,11,12] For human breast and cervical cancer cell membranes, it was found that the amount of sphingomyelin (SM) is diminished among the constituent lipids, such that the mobility is increased.[13] For a lipid membrane, three phases can generally be discerned which differ from each other by lipid packing and organisation, yielding differences in lateral diffusion (D): liquid disordered (Ld), liquid ordered (Lo) and solid gel (So) phases.[14] These phases are determined mainly by lipid composition and temperature.[15]

One of the most common techniques for phase recognition consists of the insertion of a fluorescent molecule, acting as a probe, into the lipid membrane. Several lipid membrane systems can be constructed as prototypical lipid bilayers, such as Small Unilamellar Vesicles (SUV), Large Unilamellar Vesicles (LUV) and Giant Unilamellar Vesicles (GUV).[16] When a beam of polarized light falls onto the vesicle, the probability of probe excitation depends on the orientation of its absorption transition dipole moment with respect to the electric field of the incident light, i.e. the process of photoselection.[17] The change in polarization of the fluorescence light with respect to the incident light depends on the angle between the transition moments of absorption and emission, and on the rotational mobility of the probe within the lipid membrane during the fluorescence lifetime.[18,19] Consequently, the high rotational mobility gives rise to a large depolarization. Conversely, a severely restricted rotational mobility preserves to a large extend the polarization set by the emission transition moment.[17,20,21] Hence, any change in the orientation of the transition dipole moment during the lifetime of the excited state of the probe will be responsible for a partial or total depolarization of fluorescence. From the experimental point of view, it can be remarked that also the relationships between the intensities obtained at the blue and red sides of the probe's emission spectra can be used to recognize membrane phases: the Generalized Polarization (GP) function is defined as the fraction of the differences of both emission intensities divided by their sum. It contains information about solvent dipolar relaxation processes which occur during the time the probe is excited.[22,23,24,25]

Because of this photoselective effect, several probes have been synthetized to preferentially target different membrane phases.[26,27,28] The main goal of that research is to understand or even design probes which can be versatile in the characterization of different membrane phases, i.e., they should exhibit different orientations of the transition dipole moment vector, in turn exhibiting different optical responses, with respect to different membrane phases, namely Ld, Lo and So, and therefore at a later stage report upon changing properties once the probes are applied to clinically important biological environments. An important evolution is the use of non-linear optics. To decrease the damage caused by harmful UV photons, which are mainly used in the case of an analysis based on linear absorption and conjugated probes, the use of near-IR or IR light is envisaged by means of two-photon absorption microscopy. This process involves the simultaneous absorption of two photons with lower energy to excite the molecule. This method has tremendous advantages when applied to living organisms: the rate of absorption of light by a molecule depends here on the square of the light's intensity, which is only attained in the focal point of the beam. This set-up enables an in-depth analysis and a high 3D imaging resolution, causing little or no damage to surrounding tissues. In contrast to two-photon absorption, second harmonic generation describes the scattering of photons at twice the excitation energy. Apart from the application in biological systems like fibrillary biomaterials or collagen in tissues,[29,30] second harmonic generation is used to investigate stressed or deformed systems and solid state materials by means of their quadrupole scattering.[31,32] Surface phenomena and plasmons are investigated like in the case of silver particles[33] and hemicyanine monolayers,[34] as well as the lowering of the symmetry of crystals.[35]

In previous works, we studied the 1,1'-dioctadecyl-3,3,3',3'-tetramethylindocarbocyanine (DiI-$C_{18}$(5), hereafter called DiD) probe, whose emission and absorption transition state dipole moment coincide with its long axis.[9,36] In the first one, we employed DiD to experimentally benchmark a new Bodipy-based probe and realized therefore that DiD's preference for the binary membrane mixture of Sphingomyelin (SM) and Cholesterol (Chol) in a 2:1 ratio is still unknown.[9] In the second study, we performed Molecular Dynamics (MD) simulations of DiD embedded in the biological environments of ref. [9] and focussed therefore on 1,2-Dioleoyl-sn-glycero-3-phosphocholine (DOPC) and 1,2-Dipalmitoyl-sn-glycero-3-phosphocholine (DPPC) membrane (at 323K) models, which both account for the Ld phase;[37] the binary mixture of 2:1 SM/Chol which is considered to model the Lo phase;[37,38] and the DPPC membrane model (at 298K) which mimics the So phase.[37] The calculations were analysed and the location, orientation and fluorescence anisotropy of DiD were discussed.[36] Of particular importance, the long axis of DiD is oriented almost parallel to the membrane surface. As a consequence, the DiD probe is used to visualize different parts of the unilamellar vesicles than e.g. those probes whose transition state dipole moments are found perpendicular to the membrane.[18,39,40,41] In the analysis of the depolarization after excitation of DiD, more degrees of freedom have therefore to be taken into account. Throughout the analysis, the position of this probe in the Lo SM/Chol (2:1) membrane was of particular importance as the dye was found in between the head groups of the lipids. This result is significant for a probe like DiD, which is mainly known as an Ld marker, but which had never been investigated for its role in an Lo membrane with Cholesterol in this 2:1 ratio.[28,42,43,44,45] Finally, for a mixture of DOPC (Ld) and SM/Chol (2:1, Lo), our calculated Gibbs free energy profiles provided strong indications that the DiD probe prefers to partition in the Lo phase rather than in the Ld one.

In the current work, we investigate the linear (one photon absorption-OPA) and non-linear (two photon absorption-TPA and second harmonic generation -SHG) optical as well as the fluorescence properties of the same DiD probe as the one used in our previous study.[36] The same membrane phases are studied and extensive hybrid Quantum Mechanics / Molecular Mechanics (QM/MM) calculations are performed.

**METHODS**

To properly assess the optical properties of the probe, the excited states of DiD have been investigated using Time Dependent Density Functional Theory (TDDFT) considering three different long-range corrected functionals (CAM-B3LYP, LRC-ωPBE and ωB97x-D). The calculations have been benchmarked against high-level ab-initio ones, using the Algebraic Diagrammatic Construction (ADC) scheme of the second order in both strict (ADC(2)-s) and extended (ADC(2)-x) variants, which allows the description of both singly and doubly occupied excited states. For these calculations, the 6-31G(d) Pople's basis set[46] was used and the analysis was performed by the QChem 4.3 software.[47] In view of its computational cost, only the smaller 6-31G basis set is used to obtain insight at the ADC(3) level of theory. The geometries have been optimized by means of second order Møller Plesset theory.

Forty uncorrelated snapshots of DiD in each four membranes were extracted from our previous MD simulations, as one fluorophore has been inserted into a membrane of 64 lipids per leaflet surrounded by 4500 water molecules which were described by the extended single point charge (SPC/E) model. Preparation of the membranes has been specified in detail elsewhere.[36] A box of 128 lipids in periodic boundary conditions has previously been found to be largely sufficient to prevent an interaction of the DiD dye with itself[48,49,50,51] and an analogous routine has been found to thoroughly describe and explain the experimental UV/VIS spectra of Vitamins E, C and Quercetin in a DOPC membrane.[52] Na$^+$ and Cl$^-$ ions were added to bulk water at a physiological concentration (0.9%).[36] A cylindrical 2 nm cutoff for the membrane and a semispherical 1.5 nm cutoff for the solvent and ions, centered on the DiD molecule and parallel to the z-axis perpendicular to the membrane surface, were considered to obtain the input structures for the QM/MM calculations within the electrostatic embedding method as implemented in the Dalton2016 package of programs.[53,54] In addition, the same number of snapshots were extracted from a 50 ns long MD simulation of the probe in ethanol whilst considering a spherical 2 nm cut-off centered on the probe (for this system, DiD was considered without the long alkyl chains). Within this QM/MM method, the system is split into two parts: the probe is described at the DFT level of theory through the CAM-B3LYP functional[55] and Dunning's cc-pVDZ basis set,[56] while the environment (membrane, ions and solvent molecules) is subjected to MD simulations using the GROMOS force field.[57] Hydrogen atoms have been added using the openbabel package[58,59] while the pH-level has been adjusted. The results for the snapshots have been visually verified. Due to the size of the probe (more than 160 atoms), only the first three excited states were considered in the calculations of optical properties, either in the linear (OPA) or non-linear (TPA and SHG) regimes.

The TPA ability of a system is quantified by the two-photon cross section (in GM units), computed as:[60]

$$\sigma(\omega)_{GM} = \frac{8\pi^2 \alpha^2 a_0^4 t_0}{\Gamma} \delta(\omega) \left(\frac{\omega}{2}\right)^2 \qquad (1)$$

where $\alpha$ is the fine structure constant, $a_0$ the Bohr radius, $t_0$ the atomic unit for time, $\Gamma$ the Lorentzian broadening (here considered constant with a value of 0.1 eV), $\omega$ the OPA excitation energy and $\delta$ the TPA strength in au.

Second-order non-linear optical properties of materials are assessed by the electric field induced second harmonic generation (EFISH) technique. In general, EFISH can be described as a third-order process characterized by a non-linear polarization $P^{2\omega}$. In the simple case of an optical field parallel to the applied static electric field, the electric field dependent susceptibility is reduced to:[61,62,63]

$$\chi^{(2)}(E^0) \sim \left(\langle\gamma\rangle + \frac{\bar{\beta}\cdot\bar{\mu}}{5kT}\right)^2 E_0, \qquad (2)$$

where $\langle\gamma\rangle$ is the orientationally averaged second hyperpolarizability and $\beta$ is the first hyperpolarizability. The scalar product $(\bar{\beta}\cdot\bar{\mu})$ defines the vector component projected onto the axis of the ground state dipole moment:

$$(\bar{\beta}\cdot\bar{\mu}) = \frac{1}{3}\sum_{i,j}(2\beta_{iij}\mu_j + \beta_{ijj}\mu_i). \qquad (3)$$

Since, in general, the second-order hyperpolarizability is at least one order of magnitude smaller than the first-order one, in the following analysis we will focus only on $(\bar{\beta}\cdot\bar{\mu})$, denoted as $\beta_{vec}$.

Another experimental observable related to the hyperpolarizability $\beta$ is the Hyper-Rayleigh Scattering (HRS). This scattering signal is measured by unpolarized optical excitation and can be derived as $\langle\beta_{HRS}\rangle = \sqrt{\langle\beta_{ZZZ}^2\rangle + \langle\beta_{XZZ}^2\rangle}$, where the brackets denote the orientational distribution average of the molecule in the environment. These averages can be expressed as a combination of $\beta_{ijk}$ tensor components in the frame of their molecular coordinates:[64]

$$\langle\beta_{ZZZ}^2\rangle = \frac{1}{7}\sum_i \beta_{iii}^2 + \frac{9}{35}\sum_{i\neq j}\beta_{iij}^2 + \frac{6}{35}\sum_{i\neq j}\beta_{iii}\beta_{ijj} + \frac{6}{35}\left[(\beta_{iij}\beta_{jkk} + \beta_{iik}\beta_{kjj} + \beta_{jji}\beta_{ikk}) + \frac{12}{35}\beta_{ijk}^2\right] \qquad (4)$$

$$\langle\beta_{XZZ}^2\rangle = \frac{1}{35}\sum_i \beta_{iii}^2 + \frac{11}{105}\sum_{i\neq j}\beta_{iij}^2 - \frac{2}{105}\sum_{i\neq j}\beta_{iii}\beta_{ijj} - \frac{2}{105}\left[(\beta_{iij}\beta_{jkk} + \beta_{iik}\beta_{kjj} + \beta_{jji}\beta_{ikk}) + \frac{8}{35}\beta_{ijk}^2\right] \qquad (5)$$

We would like to stress here that by construction $β_{vec}$ depends on the ground state dipole moment of the probe, while $β_{HRS}$ does not exhibit the same dependency. However, since the optically active core of DiD is fully conjugated, the major contribution of $β_{HRS}$ lies along the long molecular axis of the probe.[65]

**RESULTS AND DISCUSSIONS**

**QM/MM benchmarking and one-photon absorption**

The averaged computed absorption spectra of DiD in each of the four different membranes over the 40 snapshots are reported in Figure 1. To evaluate the effect of an isotropic environment, the absorption spectrum was also computed in ethanol. To take the solvent (ethanol) into account, DiD was put in a solvent box and a 50 ns MD simulation was run. As with the membranes, 40 uncorrelated snapshots were extracted to perform QM/MM calculations. After a profound benchmarking based on the lowest excited states (see Supplementary Information), the CAM-B3LYP functional has been selected to perform further optical analyses.

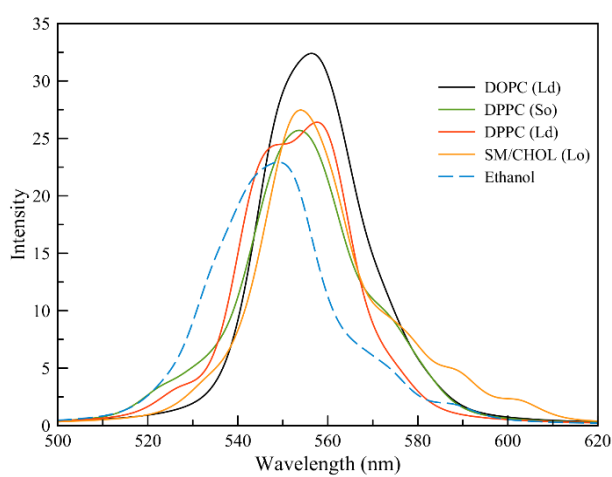

**Figure 1.** CAM-B3LYP One Photon Absorption (OPA) spectra of DiD in the different membranes: DOPC (Ld) black line, DPPC (Ld) red line, DPPC (So) green line and SM/Chol

orange line. The DiD absorption spectrum in ethanol is reported as a dotted cyan line. The absorption spectra were generated for each snapshot with a Voigt broadening of 12 nm. The resulting sum was then averaged over the 40 snapshots.

In the low energy region of the spectrum, only the first excited state S1 is bright, and this excited state experiences a red shift when an environment is present. The calculated absorption spectra of DiD in ethanol, taken as reference, can be compared with the experimental spectrum of another, closely related carbocyanine dye (DiI-C18(3)), which shows double absorption maxima at around 520 and 550 nm.[66] Apart from the different nature of the employed probes, the rigid shift observed between experimental and theoretical spectra might be due to the limitation of the computational methods. On the other hand, different shifts are observed for the absorption spectra of DiD in the various environments which can be interpreted as a consequence of the change in relative position of the dye in the membranes. Interestingly, the absorption peak maxima for DiD in membranes are only slightly red-shifted compared to ethanol (see Figure 1), which may reflect similarity in environments, mainly related to electrostatic contributions. Indeed, from our calculations,[36] the DiD head tends to partition in close contact with the polar head group region where dielectric constants are expected to be high. Conversely, the long alkyl chains of DiD are buried deeper into the bilayer. However, since the frontier orbitals are mainly localized along the conjugated backbone, the tails are not expected to directly affect its spectral properties. The maxima of the S0 → S1 transition bands are seen at 558 nm, 553 nm, 554 nm and 560 nm in DOPC, DPPC (Ld), DPPC (So) and in the SM/Chol mixture, respectively, in agreement with the value of the absorption maxima at 550 nm from experimental measurements observed in both Ld and Lo phases.[67,68] Slight differences in energy shifts were thus observed within a 10 nm range. Yet, fine structure elements are seen in the absorption spectra obtained in the different membranes; while for the case of DOPC only one broad absorption peak is present, clear shoulders can be remarked for the So and Lo phases, arising from the different local environments around the probe head. In all environments, the so-called Λ-overlap parameter for the S1 state as described by Peach et al.[69] presents a value higher than 0.64, assuring the local character of the HOMO to LUMO transition (see Figure S1 in Supplementary Information).

Considering the absorption window for each membrane, the width of the main absorption peak can be related to the rotational freedom of the probe and the subsequent local variation in the environment. Since the ethanol solvent gives rise to an isotropic environment (the probe can freely rotate in all directions), the corresponding absorption window of DiD, which amounts to 64 nm, is considered a reference value for this analysis. The presence of the anisotropic So phase of the lipid membrane (DPPC) on the other hand induces the greatest reduction of the absorption, being 40 nm in DPPC (So). This follows from the difficulty of rotational movement of the probe due to both the highly anisotropic environment and the gradient in dielectric constant along the normal of the membrane. In Ld phase (either DPPC or DOPC membranes) the absorption window is slightly greater than in DPPC (So), namely 49 and 59 nm, respectively, indicating more freedom of rotation of DiD. Finally, in the Lo phase (SM/Chol), a broader absorption window of 69 nm was observed, higher than was observed in the solvent. This particular behavior can be rationalized by the shallow penetration and the relatively large amount of water molecules around the probe in the SM/Chol bilayer, with the latter ensuring high rotational freedom.[36] It is worth mentioning that these results highlight that the combination of the fine structure of the absorption spectra and the absorption maximum might support differentiation of different membrane phases. An analysis of the structures which give rise to the shoulders seen in the here reported simulated linear absorption spectra for the different environments is given in Figures S2-S4 in Supplementary Information. To gain deeper insight into the partitioning of DiD in different membranes, non-linear optical and fluorescence analyses may become particularly instructive.

**Two-photon absorption**

The critical parameter to consider for TPA measurements and calculations is the TPA cross section σ. Interestingly, the sequence of active and inactive TPA excited states of DiD are not altered when an environment is present, compared to a vacuum. The S1 state is rather TPA forbidden in all considered environments, as the TPA cross section values range from 3 to 5 GM in the different membranes and amount to 8 GM in solvent, see Figure S5. The first active TPA state is S2, with cross section values ranging from 1300 to 1600 GM (see Figure 2), which explains the interest for DiD in *in vivo* experiments.[70,71] Despite the broad usage of this probe in phase recognition analyses, we were unable to find available TPA (and SHG) experimental data in literature. Supported by our experience and comparisons with experiments for other fluorescent probes,[72] we would like to use the computational analysis presented in the current work to predict the non-linear optical properties of DiD, to draw attention at the potential capacities of the proposed work schemes, and to eventually encourage future experiments.

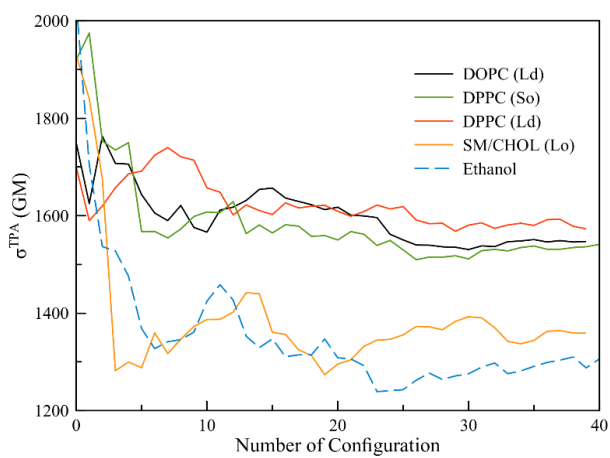

**Figure 2.** Running averaged Two Photon Absorption (TPA) cross section σ in GM units of DiD in the different membranes for the S2 excited state over the different snapshots selected from the MM calculations. DOPC (Ld) is indicated by means of a black line, DPPC (Ld) red line, DPPC (So) green line and SM/Chol orange line. The cross section in ethanol is reported as the dotted cyan line.

A thorough analysis of the S2 properties of DiD in the different environments highlighted different responses depending on the lipid phase. It should be noted that the σ value depends on the location and orientation of the transition dipole moment of the probe. In our previous study,[36] it has been found that the DiD probe is located at ~1 nm from the center of the membrane for the DPPC (Ld) and DPPC (So) membranes. For DOPC (Ld), the average distance found was between 1.3 and 1.7 nm, while for SM/Chol the probe was located just below the polar head groups, at 1.9 nm. The transition state dipole moments for the probe in SM/Chol and DPPC (So) described an angle of ~70° with the Z-axis perpendicular to the membrane, while the transition state dipole moment in the DPPC (Ld) environment amounted to 70-80°. The angle with the Z-axis for DOPC (Ld) was found to be 85°. When inserted in an isotropic environment such as ethanol, the probe experiences electrostatic interactions which are isotropic, giving rise to an averaging of the depolarization of the transition dipole moment, in turn resulting in a low σ value of 1300 GM. When DiD is inserted into a more anisotropic environment such as lipid bilayers, the differences in phases play an important role on the σ values, mainly due to the changes in charge distributions of the environment. Moreover, we have shown that the influence of the environment upon the transition state dipole moment can be quite severe, depending on the phase of the membrane and positioning of the probe.[72] In the SM/Chol binary mixture (Lo phase), where the probe is located closer to the polar head region, the depolarization is prominent, leading to a low σ value of 1350 GM. Conversely, in Ld and So phases where the probe is located deeper in the membrane[36] and is thus in closer proximity to the apolar lipid tails, the depolarization is lower and, in turn, the σ values are higher at 1540 GM, 1550 GM and 1570 GM in DPPC (So), DOPC and DPPC (Ld), respectively. Moreover, in all four membranes, the S2 state is a low-lying state which expresses the TPA absorption in a wavelength ranging from 650 to 750 nm (Figure S7).

For the brightest excited state S3, σ values are in a range between 90 000 and 120 000 GM. The highest value has been observed in the So phase (ca. 120 000 GM), followed by the Lo phase (ca. 106 000 GM) and eventually the two Ld phases (both ca. 95 000 GM). It means that the trend is reversed compared to that with the S2 excited state, namely Ld > So > Lo and So > Lo > Ld for S2 and S3, respectively (see Figure S6). Most probably, it is related to a difference in environment interactions when the two different excited states are considered, in line with our previous work.[73] As a reference, the σ values in ethanol amount to ca. 14 500 GM. The TPA for the S3 excited state is found in the 550-600 nm wavelength window (Figure S7) and has an outstanding sigma value. However, it is surpassed by the OPA events for the S1, which take place at similar wavelengths. Therefore, the first bright S2 state of DiD should be considered for TPA microscopy, owing to its low-lying TPA wavelengths and variability in the different membrane phases.

**Second harmonic generation**

Second harmonic generation (SHG) is another important non-linear optical property of a probe that can be utilized for membrane phase recognition. To be active as a SHG probe, the molecular centrosymmetry has to be disrupted. By performing molecular dynamics calculations in the NPT ensemble for DiD, fluctuations in the conformation take place, which manifest themselves also in the snapshots which are selected from the MD runs throughout the current work. For this analysis, the critical parameter is the hyperpolarizability β. As discussed in the Methods section, we consider here two different routes to compute β, which will provide complementary results for DiD in the different membranes (Figure 3).

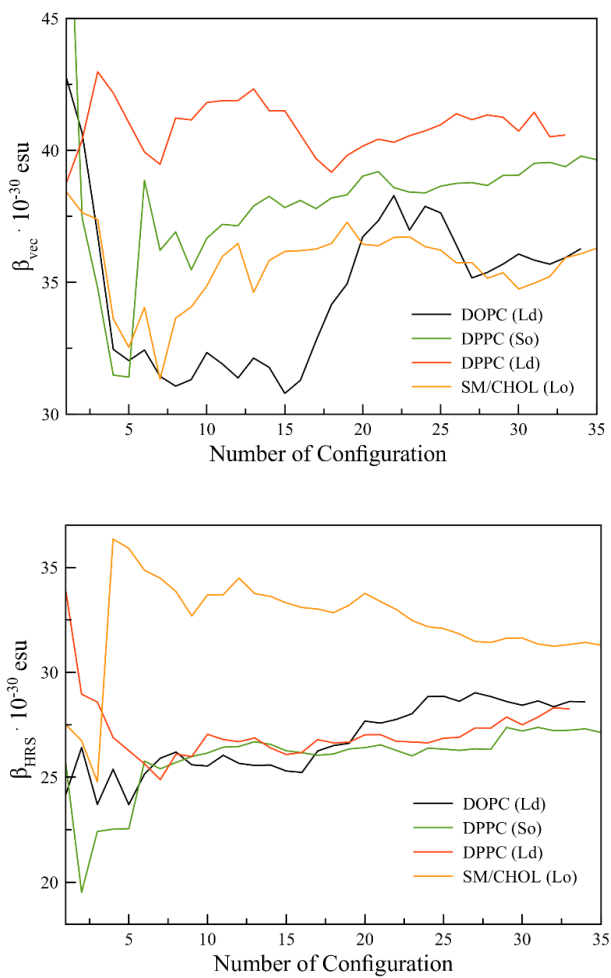

**Figure 3.** Running average for the Second Harmonic Generation (SHG) analysis in esu units. Two routes to the first hyperpolarizability of DiD in the different membranes are considered: $\beta_{vec}$ (*top*) and $\beta_{HRS}$ (*bottom*).

$\beta_{vec}$ values range from 35 to $41 \cdot 10^{-30}$ esu, depending on the membrane phase. The highest value of $41 \cdot 10^{-30}$ esu is found for DPPC (Ld), followed by $38 \cdot 10^{-30}$ esu for DPPC (So), and finally the lowest value of $35 \cdot 10^{-30}$ esu and $36 \cdot 10^{-30}$ esu was found for DOPC and the SM/Chol mixture, respectively. As a reference, $\beta_{vec}$ was also calculated in vacuum for the DiD optically-active core (i.e. discarding long alkyl chains), providing a value of $26 \cdot 10^{-30}$ esu. On the other hand, a high $\beta_{HRS}$ value of $32 \cdot 10^{-30}$ esu was obtained in SM/Chol, followed by an intermediate value of $27 \cdot 10^{-30}$ esu in both DOPC and DPPC (Ld), and a lower value of $26 \cdot 10^{-30}$ esu for DPPC (So). Again as a reference in vacuum, $\beta_{HRS}$ equals $16 \cdot 10^{-30}$ esu. From the results reported in Table 1, we observe different trends for the two β's, namely DPPC (Ld) > DPPC (So) > DOPC (Ld) ~ SM/Chol (Lo) and SM/Chol (Lo) > DOPC ~ DPPC (Ld) > DPPC (So) for $\beta_{vec}$ and $\beta_{HRS}$, respectively.

**Table 1.** Hyperpolarizabilities values for the DiD probe in vacuum and in the different membranes. All values are reported in esu. The standard deviation is also reported.

|  | Vacuum | DOPC (Ld) | DPPC (Ld) | DPPC (So) | SM/Chol (Lo) |
|---|---|---|---|---|---|
| $\beta_{vec}$ ($10^{-30}$ esu) | 26.09 | 34.76 ± 2.91 | 40.88 ± 0.91 | 38.46 ± 2.98 | 35.66 ± 1.36 |
| $\beta_{HRS}$ ($10^{-30}$ esu) | 15.67 | 26.80 ± 1.61 | 27.14 ± 1.45 | 26.09 ± 1.57 | 32.27 ± 2.21 |

The first interesting result from this SHG analysis is that the presence of the environment increases both $\beta_{vec}$ and $\beta_{HRS}$ values, and that values of $\beta_{vec}$ are always higher than the $\beta_{HRS}$ values when the same bilayer is considered. This difference can be related to the computation of the two β quantities; $\beta_{vec}$ is obtained as a projection on the dipole moment of DiD (which is perpendicular to the long molecular axis, see Figure S8), while the major contribution of $\beta_{HRS}$ lies along the molecular axis (same direction as the transition dipole moment). Moreover, $\beta_{HRS}$ does not account for the dipole, hence these values are not affected by the depolarization of the different environments, as is for $\beta_{vec}$. Going from $\beta_{vec}$ to $\beta_{HRS}$, the percentage of attenuation is 30, 21, 32 and 14 % in DPPC (Ld), DOPC, DPPC (So) and SM/Chol (Lo) phases, respectively, while in vacuum a 40% decrease has been found. Hence, there is a clear effect of the different phases in attenuation of the β values, due to the differences in polarization that the probe experiences in the different environments. This can be seen more clearly for the Lo phase, where the lowest difference between the two β values has been observed; the averaged distance between charges of the surrounding molecules and the probe in SM/Chol are greater due to the cavity (filled with water) around the head group of the probe – and the distribution of charges is clearly different as the probe is closer to the polar head of the membrane and more water molecules are found surrounding the probe. This is another manifestation of what we have seen in the case of the order parameters and single-photon induced fluorescence anisotropy, depicted by a strongly pronounced decay of the fluorescence anisotropy compared to the cases of the three other bilayers.[36] On the other hand, in the So phase, the difference of 32% between the two β values arises from the opposite behavior of the probe, since now it is located in the apolar region of the system. In addition, the charges of the molecules in the environment might influence the ground state dipole moment and consequently $\beta_{vec}$, whilst it does not have relevant influence on the large geometrical axis of the probe, resulting in lower values of $\beta_{HRS}$.[74]

**Fluorescence decay time**

In addition to the (non)-linear optical properties, the fluorescence of the probe in the different membranes can be computed from the extracted snapshots by means of a radiative decay time analysis. The fluorescence lifetime might be affected by three factors when a probe is embedded in an environment: quenching, changes in conformation of the fluorophore, and fluctuation in dielectric properties of the probe surrounding molecules. Since it is not possible to compute the quantum yield, we assume a unitary value. This means that non-radiative decay channels are neglected in the current calculations of the fluorescence decay time (no experimental data available).

The electronic properties of DiD are dominated by the delocalized π electrons of the probe head, and the fluorescence properties can be computed by reverting to the inherent radiative lifetime $\tau_0$ of the probe, which depends on the spontaneous emission $\Gamma_0$ rate through

$$\tau_0 = 1/\Gamma_0 \qquad (6)$$

and can be expanded as[75]

$$\Gamma_0 = \frac{4}{3} \frac{|\mu_{eg}|^2}{4\pi\varepsilon_0 \hbar} \left(\frac{\omega_{eg}}{c}\right)^3 \qquad (7)$$

where $\mu_{eg}$ and $\omega_{eg}$ are respectively the transition dipole moment and the transition frequency for the first one-photon allowed excited state $e$ of the probe relative to the ground state $g$; $\varepsilon_0$ denotes the permittivity of vacuum, $\hbar$ the reduced Planck constant and $c$ the light speed constant. When inserted into any environment (solvent or biological), the dielectric properties of the medium strongly alter this intrinsic lifetime (dielectric properties of the surrounding molecules). In this view, the energy of the emitted photon is renormalized through the $\varepsilon_0 \to \varepsilon_r \varepsilon_0$ and $c \to c/n$ substitutions, with $\varepsilon_r$ the relative permittivity and $n$ the refractive index of the medium. The spontaneous emission rate can now be rewritten as $\Gamma_r \to n\Gamma_0$, with $n = \sqrt{\varepsilon_r}$. In the specific case of DiD, the absorption and emission transition dipole moments for the first excited state $e$ are parallel to each other and are oriented along the long axis of the molecule. To analyze the effect of the different environments on DiD fluorescence properties, different refractive indices were considered as obtained from experiments on either supported or suspended lipid bilayers, namely 1.378 for DOPC and 1.789 for DPPC (So),[76,77] 1.438 for DPPC (Ld)[78] and 1.555 for the 2:1 SM/Chol mixture.[79] In addition, we consider also the decay time in ethanol ($n$ = 1.362) as reference for the effect of isotropic media. Subsequently, to have an orientational average distribution of the decay time, variations in probe conformations in the different environments with time were considered. In this way, instantaneous changes in position and orientation of the probe in the membranes can be analyzed. This method was applied for the first time in our work on a newly designed azobenzene derivative embedded in different membrane phases.[73] A direct comparison with the experimental FLIM technique is however not possible as only natural lifetimes are considered while non-radiative decay processes might be of importance in experiments. The results reported in Figure 4 depict differences in decay time (for the S1 excited state) in the order of tenths to 2-3 nanoseconds for the different environments.

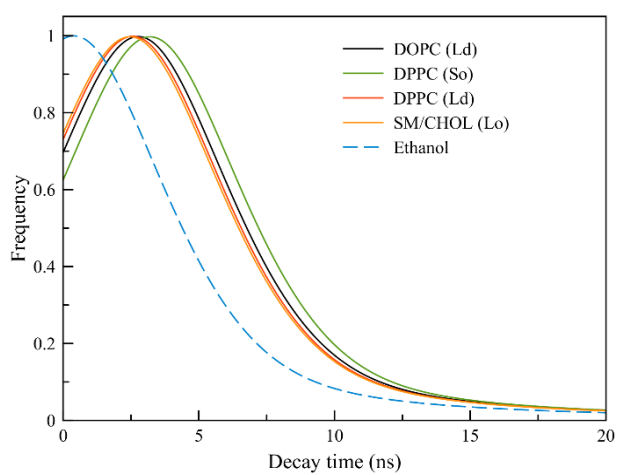

**Figure 4.** Fluorescence decay time for DiD in the different membrane phases and in ethanol. The different curves are normalized to the number of frames considered.

From Figure 4, it is possible to discriminate between very short decay times, of the order of tenths of nanoseconds for the isotropic solvent to nanosecond decay when the probe is inserted into different membranes. In particular, in ethanol DiD peaks at 0.393 ns, very similar to the experimental value of 0.22 ns observed for the smaller DiI-C18(3) compound.[80] Moreover, the peak maxima in the different membranes are found at 2.768, 2.562, 3.202 and 2.460 ns for DiD in DOPC, DPPC (Ld), DPPC (So) and SM/Chol, respectively. Hence, as experimentally measured, the same decay time trend is observed here, with So > Ld ~ Lo. These values can be qualitatively compared with a FLIM analysis done in Ld and So phases of model membranes, with decay times reported in a range 0.80-0.93 and 1.13-1.31 ns, respectively.[80,81,82,83] Although these calculated values are higher in absolute value with respect to the experimental ones, their ratio (Ld/So) is similar, being 1.41 from experiment and 1.16 from our calculations, thus ensuring a qualitative agreement with our model and the experiment. Considering the strong approximations made, this results are encouraging. In the current study, we are thus able to obtain well-separated decay time curves for DiD, which has different orientations and positions in the different membranes. The differences between lipid bilayers and solvent environments on decay time can be finally addressed; when embedded in membranes, DiD experiences different local interactions which depend on its position and orientation as well as on the local heterogeneity of the environment. A fine balance between charge (de)polarization, viscosity of the environment and interactions should be considered for this analysis, since they can affect each other in the determination of the decay time. The stiffness of DPPC in the So phase and the strong interaction with the membrane tails lead to the longest decay time calculated among the different membranes, while in both lipids representing the Ld phase this value is significantly lower, due to the less tight packing of the molecules and therefore a reduced inductive effect of the probe on the neighboring lipids.[3] A clear case where the interaction plays the most important role is the SM/Chol mixture; in this membrane, DiD is located very close to the head of the bilayer, experiencing an enhanced influence of both the polar groups and the relatively high amount of water molecules which results in a similar decay time as in the Ld phase, which agrees with the results by Margineanu *et al.* who investigated a perylene derivative in membranes in Ld and Lo phases.[3] We would like to add that our results are obtained for a Lo phase in which only SM and Chol are considered, omitting the influence of a minimal contribution of e.g. DOPC lipids which might be present in a ternary DOPC/SM/Chol mixture.[84]

**CONCLUSIONS**

In this work, the optical properties of DiD embedded in different lipid bilayers have been studied by means of computational simulations. From QM/MM calculations and subsequent analyses, we found that both OPA and TPA properties of DiD may discriminate the different membrane phases. In particular, OPA spectra exhibit different fine structures and provide together with the related absorption windows useful criteria to discriminate between the different phases. From the TPA analysis, we observed very high cross section values, in the order of thousands of GM, with strong differences between the membrane phases, with the following trend: Ld > So > Lo. In addition, the SHG properties for DiD were analyzed: the hyperpolarizabilities show different trends, namely DPPC (Ld) > DPPC (So) > DOPC ~ SM/Chol for $\beta_{vec}$ and SM/Chol > DOPC ~ DPPC (Ld) > DPPC (So) for $\beta_{HRS}$. Moreover, the strength of this probe is enhanced by the rotational averaged distribution of the fluorescence decay time. From this analysis, we observe a different trend with decay time So > Ld ~ Lo. Thus, the combination of these OPA, TPA, SHG and fluorescence decay analyses provides a comprehensive toolbox exploiting the full potential of DiD as universal probe for lipid phase recognition.

**Acknowledgements**

S.O. is grateful to the Center for Quantum Materials and Nordita for his funding in Sweden. S.O. acknowledges the National Science Centre, Poland, grant UMO-2015/19/P/ST4/03636 for the funding from the European Union's Horizon 2020 research and innovation program under the Marie Skłodowska-Curie grant agreement No. 665778. The authors thank the Swedish Infrastructure Committee (SNIC) for the computational time granted within the medium allocations in 2016 (1-87, 1-415 and 1-465) and 2017 (1-16, 1-102), as well as the small ones 2015/4-44 and 2017/5-3. The authors are also grateful to the "Consortium des Équipements de Calcul Intensif" (Céci, with the *Lemaitre2*, *Nic4*, *Hmem*, *Hercules* and *Dragon1* clusters) of the Association Wallonie-Bruxelles. The Flemish Supercomputer Centre (VSC) (Flanders, Belgium) and the Herculesstichting (Flanders, Belgium) are acknowledged for the generously allocated computational time on the Tier-1 cluster *Breniac* as well as the Tier-2 cluster *Thinking*. P.T. thanks the Czech Science Foundation (P208/12/G016) and National Program of Sustainability I from the Ministry-of-Youth, Education and Sports of the Czech Republic (LO1305). The authors are grateful to Lindsay Leach for language consultation.

**Supporting Information**

Details of the benchmark and two-photon absorption spectra of the probe in the different membranes are reported. This information is available free of charge via the Internet at http://pubs.acs.org.

**TOC:**

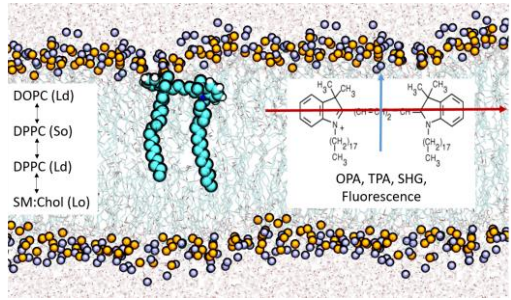